\documentclass[aps,prb,reprint,superscriptaddress]{revtex4-1}
\usepackage{amsmath}
\usepackage{multirow}
\usepackage{graphicx}
\usepackage{epstopdf}
\usepackage{tabularx, booktabs}
\usepackage{pbox}
\newcolumntype{Y}{>{\centering\arraybackslash}X}
\usepackage[usenames,dvipsnames]{color}
\usepackage[caption=false,position=b,singlelinecheck=off,font=normalsize,labelfont=bf,justification=justified]{subfig}
\usepackage{hyperref}

\hypersetup{
	colorlinks=true,
	linkcolor=blue,
	filecolor=black,      
	urlcolor=blue,
	citecolor=blue,
}
\graphicspath{ {Figures/} }

\newcommand{\Na}{Na$_2$RuO$_3$}
\newcommand{\NaO}{Na$_2$RuO$_4$}

\newcommand{\CaO}{Ca$_2$RuO$_4$}

\begin{document}

\title{Correlated electron metal properties of the honeycomb ruthenate Na$_2$RuO$_3$ }

\author{L.S.I. Veiga}
\affiliation{London Centre for Nanotechnology and Department of Physics and Astronomy, University College London, Gower Street, London WC1E 6BT, UK}
\affiliation{Diamond Light Source Ltd., Harwell Science \& Innovation Campus, Didcot, Oxfordshire OX11 0DE, UK} 

\author{M. Etter}
\affiliation{Deutsches Elektronen-Synchrotron (DESY), Hamburg 22607, Germany}

\author{E. Cappelli}
\affiliation{Department of Quantum Matter Physics, University of Geneva, 24 Quai Ernest-Ansermet, 1211 Geneva 4, Switzerland}

\author{H. Jacobsen}
\affiliation{Clarendon Laboratory, Department of Physics, University of Oxford, Oxford OX1 3PU, UK}
\affiliation{Paul Scherrer Institute, Laboratory for Neutron Scattering and Imaging, 5232 Villigen, Switzerland}

\author{J.G. Vale}
\affiliation{London Centre for Nanotechnology and Department of Physics and Astronomy, University College London, Gower Street, London WC1E 6BT, UK}

\author{C. D. Dashwood}
\affiliation{London Centre for Nanotechnology and Department of Physics and Astronomy, University College London, Gower Street, London WC1E 6BT, UK}

\author{D. Le}
\affiliation{ISIS Facility, Rutherford Appleton Laboratory, STFC, Chilton, Didcot, OX11 0QX, United Kingdom}

\author{F. Baumberger}
\affiliation{Department of Quantum Matter Physics, University of Geneva, 24 Quai Ernest-Ansermet, 1211 Geneva 4, Switzerland}
\affiliation{Swiss Light Source, Paul Scherrer Institute, CH-5232 Villigen, Switzerland}

\author{D.F. McMorrow}
\affiliation{London Centre for Nanotechnology and Department of Physics and Astronomy, University College London, Gower Street, London WC1E 6BT, UK}

\author{R.S. Perry}
\affiliation{London Centre for Nanotechnology and Institute for Materials Discovery, University College London, Gower Street, London WC1E 6BT, UK}

\date{\today}

\begin{abstract}
We report the synthesis and characterisation of polycrystalline {\Na}, a layered material in which the Ru$^{4+}$ ($4d^4$ configuration) form a honeycomb lattice. The optimal synthesis condition was found to produce a nearly ordered \Na{} ($C2/c$ phase), as assessed from the refinement of the time-of-flight neutron powder diffraction. Magnetic susceptibility measurements reveal a large temperature-independent Pauli paramagnetism ($\chi_0 \sim 1.42(2)\times10^{-3}$ emu/mol Oe) with no evidence of magnetic ordering down to 1.5 K, and with an absence of dynamic magnetic correlations, as evidenced by neutron scattering spectroscopy. The intrinsic susceptibility ($\chi_0$) together with the Sommerfeld coeficient of $\gamma=11.7(2)$ mJ/Ru mol K$^2$ estimated from heat capacity measurements, gives an enhanced Wilson ratio of $R_W\approx8.9(1)$, suggesting that magnetic correlations may be present in this material. While transport measurements on pressed pellets show nonmetallic behaviour, photoemission spectrocopy indicate a small but finite density of states at the Fermi energy, suggesting that the bulk material is metallic. Except for resistivity measurements, which may have been compromised by near surface and interface effects, all other probes indicate that {\Na} is a moderately correlated electron metal. Our results thus stand in contrast to earlier reports that \Na{} is an antiferromagnetic insulator at low temperatures.

%We report the synthesis and characterisation of polycrystalline Na2RuO3, a layered material in which the Ru4+ (4d4 configuration) form a honeycomb lattice. The optimal synthesis condition was found to produce a nearly ordered Na2RuO3 (C2/c phase), as assessed from the refinement of the time-of-flight neutron powder diffraction. Magnetic susceptibility measurements reveal a large (COMPARED to what?) temperature-independent Pauli paramagnetism with no evidence of magnetic ordering down to 1.5 K, and an absence of magnetic correlations in data from  inelastic neutron scattering experiments. While transport measurements on pressed pellets show nonmetallic behaviour, heat capacity measurements and photoemission  electron spectroscopy indicate a small but finite density of states at the Fermi energy, suggesting that the bulk material is metallic. Except for resistivity measurements, which may have been compromised by near surface and interface effects, all other probes indicate that Na2RuO3 is a moderately correlated electron metal. Our results thus stand in contrast to earlier reports that Na2RuO3 is an antiferromagnetic insulator at low temperatures.

\end{abstract}

\maketitle

\section{Introduction}
The identification and characterisation of new quantum materials remains one of the principal drivers of research into complex systems. Quantum materials exhibit a plethora of functional properties, from superconductivity to topological states, driven by the subtle interplay between competing energy scales~\cite{Keimer2017, Savary2016}. The inherent instability to perturbation in these systems is both blessing and curse; exotic ground states can be stabilised out of the competition, but these delicate states are often extremely sensitive to crystalline disorder and impurities~\cite{Balents2010, Basov2017}. Hence, robust material science research is crucial to ensure well-characterised and phase pure samples are provided for advanced measurement techniques like neutron scattering, angle-resolved photoemisison spectroscopy and scanning tunnelling microscopy ~\cite{Samarth2017}. In this paper, we report a comprehensive powder synthesis and characterisation study of the layered honeycomb quantum material {\Na}. We determine its ground state to be a correlated electron metal, which is at odds with previous claims of insulating antiferromagnetism~\cite{Wang2014}. 

%\textcolor{red}{a material of topical interest due to its potential to host rich physics from the spin-orbit coupled $J_{\rm{eff}}=0$ state. Instead, we determine its}

Layered honeycomb structures have been identified as having the potential to host exotic ground states, a well known example of which is the honeycomb iridates A$_2$IrO$_3$ (A=Na, Li). In these Ir$^{4+}$ (5$d^5$) materials, the strong spin-orbit coupling (SOC) stabilises a $J_{\rm{eff}}=1/2$ Mott insulating state. The combination of strong SOC, intrinsic geometric frustration and other competing terms in the Hamiltonian favours highly anisotropic Kitaev exchange interactions~\cite{Jackeli2009, Chaloupka2010}. Theoretical predictions suggest that the Kitaev physics associated with these interactions may lead to exotic phases such as topological insulators~\cite{Chen2015} and quantum spin liquid state~\cite{Kitaev2006, Witczak-Krempa2014} (QSL). However, such novel ground states have not yet been realized in these systems largely due to competing isotropic interactions (Heisenberg) that stabilize magnetic order and/or lattice distortions that relieve geometrical frustration.

In this context, the study of the honeycomb ruthenate counterparts (A$_2$RuO$_3$, with A=Na, Li) provides a new path to explore systems with unusual physics emerging from the interplay of SOC and electronic correlations. These materials feature a Ru$^{4+}$ (4$d^4$) configuration, which in the presence of a strong octahedral crystal field, and intermediate strength SOC and Hund's couplings,  yields a $t_{2g}^4$ ground state manifold with $J_{\rm{eff}}=0$. For such systems it has recently been proposed that super exchange interactions lead to excitonic Van Vleck-like magnetism, and above a quantum critical point, Bose Einstein condensation of the higher lying $J_{\rm{eff}}=1$ triplet occurs~\cite{Khaliullin2013, Meetei2015, Svoboda2017}. The inherent bond-directionality of the $J_{\rm{eff}}$ states, renders the ground state exquisitely sensitive to the lattice connectivity. In the case of the honeycomb lattice it is predicted that a novel type of spin-liquid may be realised~\cite{Chaloupka2019}.

\begin{figure}
	\centering
	\includegraphics[width=0.8 \columnwidth]{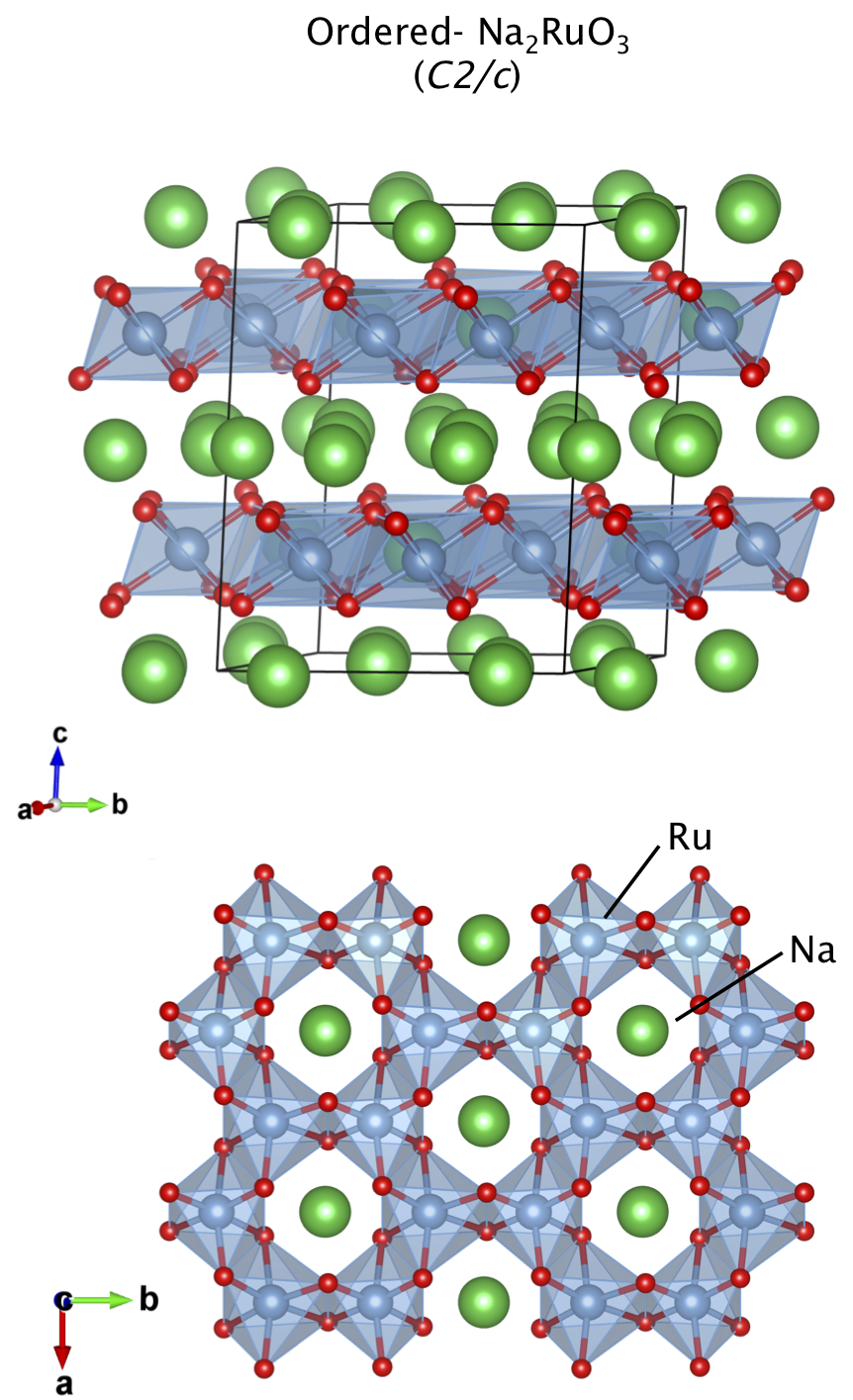}

	\caption{(Color online) Crystal structure of the ordered phase of {\Na} in the $C2/c$ space group. Oxygen ions are drawn in red, Na/Ru cations are showed in green/blue. Ordered {\Na} has the honeycomb-type cation ordering in the [Na$_{1/3}$Ru$_{2/3}$]O$_2$ slab. Stacking faults occur through the occasional shift of the [Na$_{1/3}$Ru$_{2/3}$]O$_2$ layers perpendicular to the stacking direction, which in this case is the c-axis.}
	\label{structure}

\end{figure}

While interest in the honeycomb ruthenates has been largely concentrated in the Li$_2$RuO$_3$ due to the presence of dimerization and the formation of an exotic valence bond liquid phase~\cite{Miura2007, Kimber2014, Park2016}, the studies of the physical properties of {\Na} are somewhat scarce and unclear. Early studies by Mogare et al.~\cite{Mogare2004} identified the crystal structure and synthesis conditions of {\Na}, noting that the material suffered from cation disorder within the honeycomb plane. More recently, Cao and co-workers~\cite{Wang2014} claimed to synthesise single crystals, determining {\Na} to be an insulating antiferromagnet with a N\'{e}el temperature around 30 K from bulk susceptibility and heat capacity measurements. Subsequently, Gapontsev et. al.~\cite{Gapontsev2017} used X-ray absorption spectroscopy, resistivity and density functional theory calculations to propose a model of the magnetic structure in the ground state. In parallel, {\Na} has been identified as a candidate material for a cathode for sodium batteries~\cite{Boisse2016, Boisse2019}. It has been demonstrated that it exhibits a remarkably high reversibility of electrochemical Na insertion/deinsertion combined with a high specific capacity~\cite{Tamaru2013}. 
From all these studies, {\Na} has been identified to crystalise in one of the two existing polymorphs: a disordered-{\Na} with space group $R\overline{3}m$~\cite{Tamaru2013} and randomly distributed [Na$_{1/3}$Ru$_{2/3}$]O$_2$ slabs and an ordered-{\Na}~\cite{Mogare2004,Boisse2016, Boisse2019,Wang2014}, with space group $C2/c$ (or $C2/m$) and honeycomb-ordered [Na$_{1/3}$Ru$_{2/3}$]O$_2$ slabs (see Figure~\ref{structure} for the $C2/c$ crystal structure). Our interest is in the \textit{ordered phase}, which might be expected to host exotic physics in the context of a $d^4$ spin-orbit coupled Mott insulator~\cite{Chaloupka2019}. The purpose of the current study is to revisit the synthesis and characterisation of low-temperature properties to clarify the ground state of the \textit{ordered} phase of {\Na}. Remarkably, we find no evidence of magnetism in this material, identifying it instead as a paramagnetic, correlated electron metal. We discuss the possible reasons why this material was miscategorized.

%The room temperature structure was well characterised using synchrotron radiation to determine the C2/m space group with a lamellar alpha-NaFeO2-type structure consisting of Na and Na1/3Ru2/3 layers (see figure ???). 

%The presence of superlattice peaks in the Bragg diffraction pattern between 20deg and 30deg using Cu-Kalpha radiation indicates an ordered C2/m structure.

\section{Experiment}\label{sec:experiment}

\subsection{Powder synthesis}
The synthesis of polycrystalline {\Na} consisted of two steps, following similar procedures present in Refs.~\onlinecite{Mogare2004, Shikano2004}. The first part involved the synthesis of {\NaO}, prepared by solid state reaction from stoichiometric amounts of Na$_2$O$_2$ and RuO$_2$. The powders were ground and mixed in an agate mortar under argon. The mixture were placed in an alumina crucible and heated to 450 $^\circ$C for 5 h, 530 $^\circ$C for 10 h and 630 $^\circ$C for 20 h under oxygen atmosphere with a  temperature rate of 50 $^\circ$C/h. The phase composition of the resulting powder was checked using a Rigaku Miniflex 600 x-ray powder diffractometer, with the resulting diffraction pattern showing pure {\NaO} phase. The second step consisted of reducing the {\NaO} under argon flow at 950 $^\circ$C for 24 h using a tube furnace. The resulting product was a black homogeneous {\Na} powder, which was found to be mildly air sensitive. Efforts to make single crystals via flux growth were unsuccessful, due in part to the volatility of the material and the requirement of having a reducing atmosphere to stabilise the Ru$^{4+}$ ion.

\begin{figure}
	\centering
	\includegraphics[width=1.\linewidth]{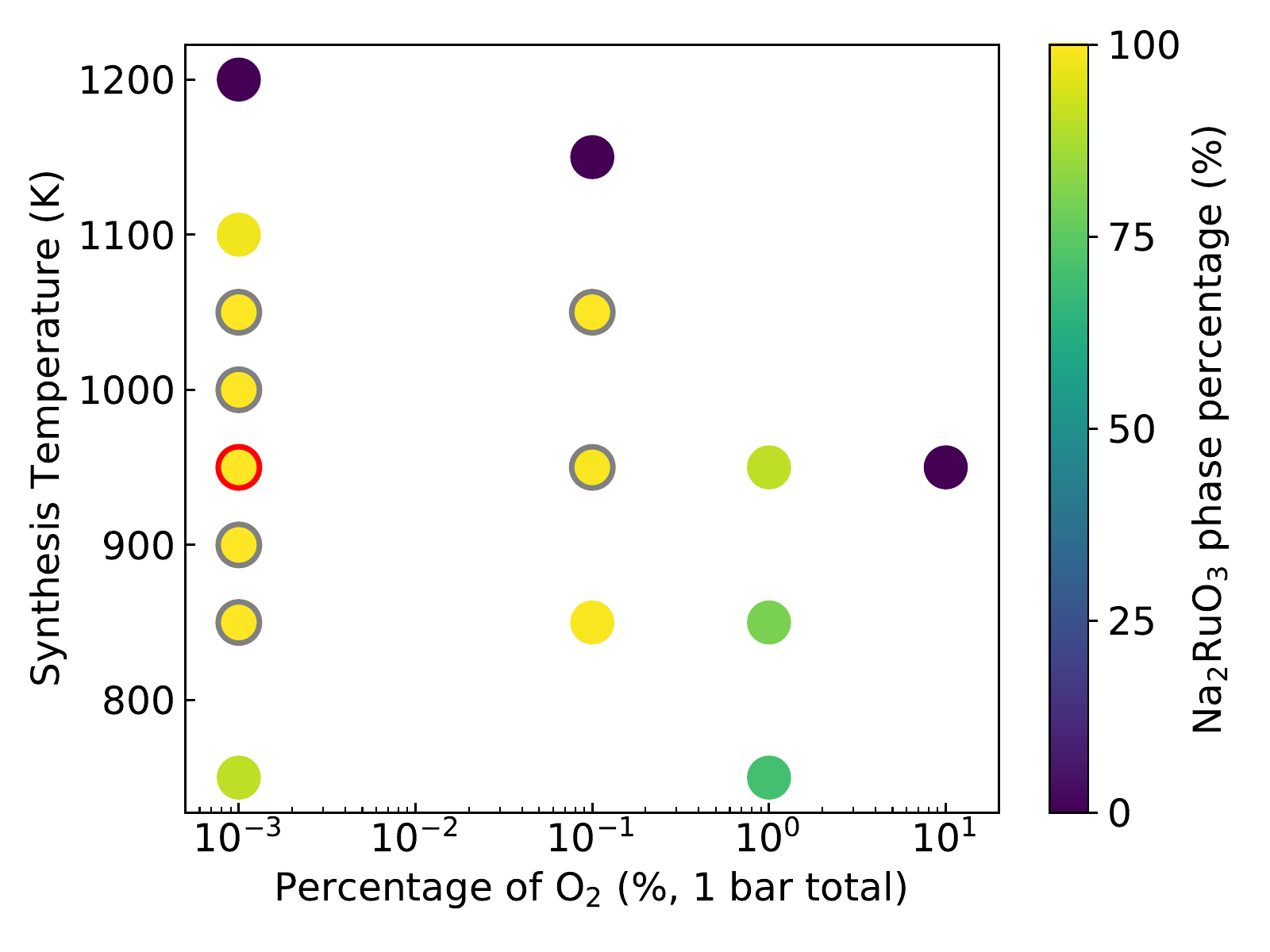}

	\caption{Phase diagram of the synthesis conditions of {\Na}. The markers denote the conditions at which the synthesis attempts were conducted. The color scale represents the estimate percentage of the ordered {\Na} phase compared to foreign phases obtained from the x-ray diffraction measurements. The foreign phases include other Na-Ru-O compositions such as disordered {\Na}, {\NaO}, NaRuO$_2$, Na$_3$RuO$_4$ and Na$_{27}$Ru$_{14}$O$_{48}$. The circles with grey edge indicate the optimum synthesis conditions. The circle with red edge highlights the synthesis conditions used to prepare the samples for all the measurements present in this manuscript.}
	\label{phase_diagram}
\end{figure}

Attempts to synthesize {\Na} from {\NaO} using other synthesis temperatures and atmospheric conditions were also conducted and are displayed in the phase diagram of Figure~\ref{phase_diagram}. The phase homogeneity of the different batches was verified by powder x-ray diffraction\footnote{See Supplemental Material at [URL will be inserted by publisher] for further information.}. In general, the optimum conditions for the synthesis of \textit{red}{ordered} {\Na} phase was found to be in the region corresponding to synthesis temperatures between 850 $^\circ$C to 1050 $^\circ$C and partial oxygen pressure in one bar varying from 0.001\% (5N argon atmosphere) to 0.1\%. In general, the final structure was sensitive to final synthesis conditions; for example, temperatures above 1050$^\circ$C under argon atmospheric condition produced samples with a high degree of in-plane Ru-Na disorder characterised by an $R\overline{3}m$ space group. On the other hand, for temperatures ranging from 850 $^\circ$C to 1050 $^\circ$C under the same conditions, the crystal structure of the powders were mostly characterised by the ordered $C2/c$ space group~\cite{Note1}. A complete systematic study of the bulk properties of all batches was not performed and is beyond the scope of this study. Note that despite ranges of temperature and partial oxygen pressure having been identified as optimum conditions for the synthesis of ordered \Na{} (see Figure~\ref{phase_diagram}), all the samples used in the various experiments reported in this manuscript were prepared by heating \NaO{} under argon flow at 950 $^\circ$C for 24 h.

%Attempts to synthesize {\Na} from {\NaO} using other temperatures and atmosphere conditions were also conducted and are described in the Supplemental Material. \textcolor{red}{}. 

%and neutron diffraction}, as discussed in the next sections

%Disordered {\Na} has randomly stacked [Na$_{1/3}$Ru$_{2/3}$]O$_2$ layers.

\subsection{Characterization methods}

%Once the optimum synthesis conditions were established and \textcolor{red}{the ordered} phase composition obtained, 

The crystallographic structure of the ordered \Na{} was verified by time-of-flight neutron powder diffraction (TOF-NPD) measurements on the WISH instrument at the ISIS spallation neutron source, UK. The powder ($\sim 3.5$ grams total mass) was finely ground and enclosed in a thin walled cylindrical vanadium can. The obtained powder diffraction patterns collected for temperatures ranging from $T=100$ K to $T=1.5$ K using detector banks 2 - 5 were analyzed by Rietveld refinement as implemented in the TOPAS software (TOPAS 6.0, Bruker AXS, 2017). For each diffraction pattern Chebyshev polynomials of 24th order were used in order to model the complex corrugated background. Structural parameters, like lattice parameters, atomic positions, occupancies, isotropic displacement parameters and sample dependent peak shape contributions were constrained over the different banks. More details about the Rietveld refinement can be found in the Supplemental Material~\cite{Note1}.

Bulk characterisation of the powders consisted of magnetic susceptibility ($\chi$) measurements in a Quantum Design superconducting quantum interference device (SQUID), heat capacity ($C_p$) and four-probe resistance ($R$)  measurements in a Quantum Design physical properties measurement system (PPMS). The heat capacity measurement was made on a sintered, cylindrical pellet with specific dimensions to adapt to the mounting on the PPMS addenda, i.e. large surface area and thin sample to enable good thermalisation. The two-tau analysis method was employed and a high coupling constant ($>$ 90\%) was measured at all temperatures demonstrating that the extracted time constant represents the relaxation of the sample with the external thermal reservoir. Furthermore, the relaxation curves were well fitted to the two-tau model and no distortion in the raw  relaxation  curves  was  observed. Energy dispersive x-ray spectroscopy (EDX) was used to investigate the Na:Ru ratio on a representative batch of powders using a JEOL JSM-6610LV with Oxford Instruments EDX. To this end, the powders were pelletized and cleaved to expose fresh surfaces and the measurements were conducted in six different areas of the sample. The measurements gave an average Na:Ru ratio of $\sim2.11$, suggesting that the samples are slightly Ru deficient. The deficiency of Ru is likely related to the loss of RuO$_2$ during the reduction process under argon atmosphere. While an excess of sodium might be present in our samples, due to the flexibility of the structure and as demonstrated studies of battery material~\cite{Boisse2016, Boisse2019}, the high volatility of Na$_2$O at 950 $^{\circ}$C compared to RuO$_2$ suggests that the excess of sodium in the structure of \Na{} is unlikely to occur.

The photoelectron spectroscopy (PES) measurements reported here were performed with a frequency converted continuous wave laser and a hemispherical electron spectrometer. The photon energy of 6.01 eV used in these experiments provides an increased bulk sensitivity as compared to the more widely used energies of 20-100 eV. The measurements were performed on pressed powders cleaved under ultra high vacuum and the data were collected for temperatures varying between 300 K and 10 K.

%in a chamber equipped with a monochromatic He-discharge lamp ($E=21.2$ eV) [include more details of the chamber and measurements].  

The inelastic neutron scattering measurements were performed at the time-of-flight chopper spectrometer MARI of the ISIS Facility with incident neutron energies of 180 meV, 15 meV at temperatures of 5 K, 7 K and 300 K. Powder of {\Na} ($\approx 10$ grams) were wrapped in a thin Al-foil and mounted inside a thin walled Al-can which was cooled down to 5 K inside a top loading closed cycle refrigerator (CCR) with He-exchange gas.

\begin{figure}
	\centering
	\includegraphics[width=1.\columnwidth]{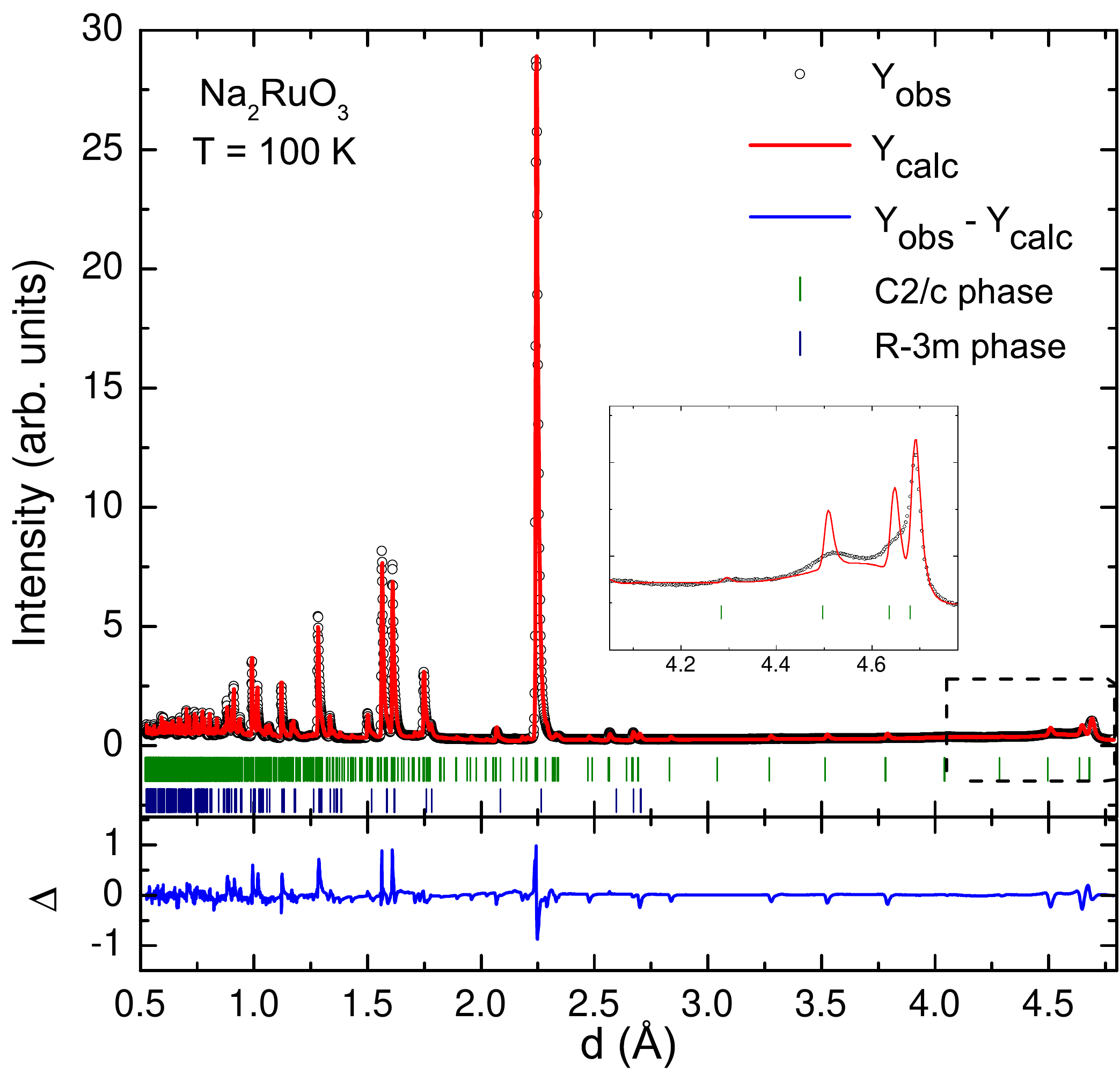}

	\caption{(Color online) Powder neutron diffraction obtained from WISH instrument at ISIS at 100 K using bank 5. Black open circles indicate the experimental data points and red solid line is the Rietveld refinement. The Bragg peak positions of the monoclinic and rhombohedral phases are indicated by the green and dark blue vertical ticks, respectively. The blue line shows the difference between the data and the simulated curve. Details of the refinement in the other detector banks can be found in the Supplemental Material~\cite{Note1}.}
	\label{NPD}

\end{figure}

\section{Results}

\subsection{Crystal structure}

\begin{table}
\caption{Final parameters of the Rietveld refinement of {\Na} using the two-phase model at $T = 100$ K. Residual factors and goodness of fit (GoF) are the overall factors for all detector banks and are defined as in the TOPAS software. The low GoF factor is explained by an overestimated R$_{\rm{exp}}$ factor. The site occupancies were kept fixed throughout the refinement.}

\begin{tabular}{ccc}
\hline
\hline 
 Formula (nominal) &   \multicolumn{2}{c}{\Na}\\
Crystal system &  monoclinic & rhombohedral\\
Space group &  $C2/c$ (N$^\circ 15$) & $R\overline{3}m$ (N$^\circ$ 166)\\
Lattice parameters &  & \\
$a$ ({\AA})         & 5.4131(1)  & 3.1689(9) \\
$b$ ({\AA})         & 9.3590(2)  & \\
$c$ ({\AA})         & 10.8113(1)  & 16.0372(55) \\
$\beta$ ($^\circ$) & 99.642(2)  & \\ 
Unit cell volume ({\AA}$^3$)  &	539.977(21)  & 139.466(94) \\
Unit cell mass (u)	   & 1560.382   & 390.095 \\
Formula units	      &  $Z=8$      & $Z=3$\\
Weight percentage     & 93.935(4)\% & 6.065(4)\% \\
R$_{\rm{exp}}$	       &  \multicolumn{2}{c}{70.508}\\
R'$_{\rm{exp}}$        & \multicolumn{2}{c}{ 75.024	}\\
R$_{\rm{wp}}$	      &  \multicolumn{2}{c}{6.885}\\
R'$_{\rm{wp}}$	      & \multicolumn{2}{c}{7.326}\\
R$_{\rm{p}}$ 	       & \multicolumn{2}{c}{9.291}\\
R'$_{\rm{p}}$ 	      & \multicolumn{2}{c}{17.820}\\
GoF	                & \multicolumn{2}{c}{0.098}\\
 
\hline
\hline

\label{tab1}
\end{tabular}
\end{table}

\begin{table}
%\squeezetable
\caption{Final refined atomic site parameters of the Rietveld refinement of {\Na} at $T = 100$ K using the two-phase model.}
\resizebox{\columnwidth}{!}{
\begin{tabular}{cccccc}
\hline
\hline 
Atom& $x$ &  $y$ & $z$ & Occupancy & B$_{\rm{iso}}$\\
\hline 
\\
monoclinic phase&&&& \\
Na1 &	0.2367(13) & 0.5956(7) & 0.0050(8) & 1 & \multirow{3}{*}{0.71(5)} \\
Na2 &	0.25 & 0.25 & 0 & 1 &  \\
Na3 &	0 & 0.8958(12) & 0.25 & 1 &  \\
Ru1 &	0 & 0.2654(7) & 0.25 & 1 & \multirow{2}{*}{0.49(4)}  \\
Ru2 &	0 & 0.5975(7) & 0.25 & 1 &  \\
O1  &	0.1295(9) & 0.0763(6) &  0.1495(5) & 1 &  \multirow{3}{*}{1.62(3)} \\
O2  &	0.1497(9) & 0.4072(6) & 0.1470(5) & 1 &  \\
O3  &	0.1554(10) & 0.7497(8) & 0.1479(6) & 1 &  \\
\\
rhombohedral phase &&&&\\
Na1 &	0 & 0 & 0 & 1 & 0.71(5) \\
Na2 &	0 & 0 & 0.5 & 1/3 & \multirow{2}{*}{ 0.49(4) }  \\
Ru &	0 & 0 & 0.5 & 2/3 & \\
O  &	0 & 0 &  0.2240(9) & 1 &  1.62(3) \\

 \hline
\hline 
\label{tab2}
\end{tabular}
}
\end{table}

\begin{figure*}
	\centering
	\includegraphics[width=2 \columnwidth]{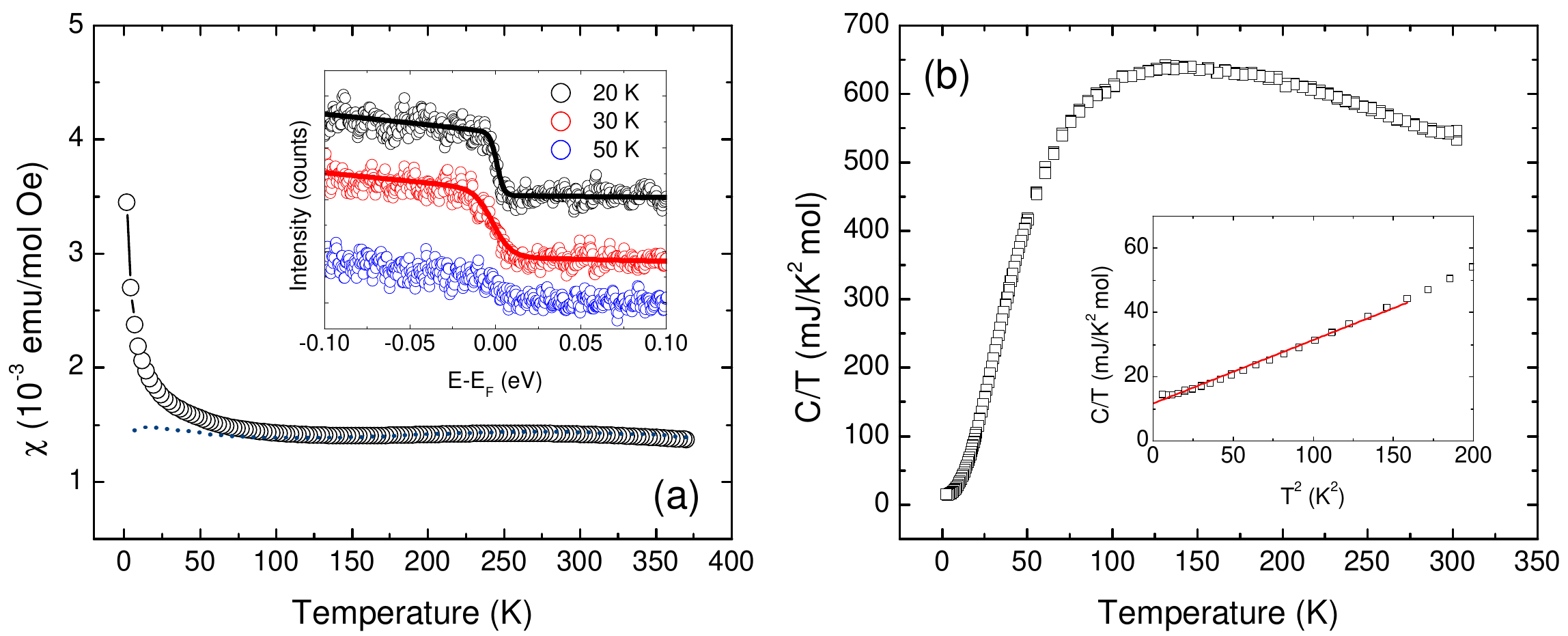}

	\caption{(Color online) (a) Magnetic susceptibility of polycrystalline {\Na} for applied field of 1 T. The raw data (circles and solid line) is shown along with the presumed intrinsic $\chi_0$ (dotted line) after subtracting a low-temperature Curie-like $1/T$ contribution. The inset shows the Fermi edge of Na$_2$RuO$_3$ polycrystalline pellets at different temperatures revealed by PES measurements (open circles). The photoemission data was fitted to a Fermi-Dirac function multiplied by a polynomial background and convoluted with a Gaussian. The Gaussian function has a FWHM of 5 meV, which is the expected resolution of the experiment. The data was stacked vertically for better visualisation of the fitting results, which are showed by solid lines. We have not fitted the data at 50 K due to its low signal-to-noise ratio. The fitting results from the 20 K and 30 K data revealed that a gap can be excluded within our experimental accuracy. (b) Specific heat data as a function of temperature. The inset shows a linear fit (solid line) in a $C/T$ vs $T^2$ plot.}
	\label{MT_HC}
%The fit curve is a polynomial background convolved with a Fermi function and a Gaussian. The Gaussian has a 5-meV FWHM, which is the expected experimental resolution of the experiment.
\end{figure*}

The TOF-NPD patterns of {\Na} were collected for temperatures ranging from $T=100$ K (Figure~\ref{NPD}) to 1.5 K~\cite{Note1}. No extra peaks were observed in the entire temperature range measured indicating no change in crystal symmetry or appearance of magnetic peaks. No secondary phases were detected. The TOF-NPD patterns were refined by the Rietveld method using existing models for {\Na} present in literature\cite{Mogare2004, Boisse2016, Boisse2019}. While the disordered crystal structure model (space group $R\overline{3}m$) fitted most of the intense reflections, several peaks were not covered by this space group, specially in the region of the inset of Figure~\ref{NPD}. The presence of these reflections demonstrate that the majority phase is the ordered $C2/c$ structure. These reflections also exhibit characteristic triangular shape profile, which is a signature of stacking fault disorder. The intrinsic stacking fault behavior arises from deviations of the theoretical optimal alternating stacking sequence of a Na layer followed by a NaRu$_2$ layer. In order to cover all the reflections present in the neutron powder pattern, we have used a two-phase model where the \textit{ordered} $C2/c$ and \textit{disordered} $R\overline{3}m$ phases were refined simultaneously. Figure~\ref{NPD} shows the Rietveld refinement of {\Na} powder neutron diffraction pattern at $T=100$ K using this approach. The refined parameters and atomic positions can be found in Tables~\ref{tab1} and \ref{tab2}. Details of the refinement at $T=1.5$ K can be found in the Supplemental Material~\cite{Note1}.    
%The first description of the crystal structure given by Morgare et. al.~\cite{Mogare2004} consists of two phases, disordered-{\Na} with space group $R\overline{3}m$ and ordered-{\Na}, with space group $C2/c$.

The two-phase model was also used by Mogare et. al.~\cite{Mogare2004} and resulted in a significant volume contribution of the disordered phase in their x-ray powder diffraction refinement, indicating a high degree of disorder in the structure of their material. In contrast, our Rietveld refinement shows a weight percentage of ordered $C2/c$ phase of $93\%$. While the two-phase model can predict most of the reflections present in the inset of Figure~\ref{NPD}, it is still not enough to account for the underlying stacking fault behavior in {\Na}, which is discussed in more detail in Section~\ref{discussion}. Moreover, the model overestimates the crystallinity of the material and creates some additional phase peaks (such as in the region $d \approx 2.5 - 4$ {\AA}) that cannot be found in the observed data.

%that our powders are constituted by a majority

%Constraining the occupancy of the Na and Ru atom on each independent crystallographic atomic site to 1 leads to a disordered model with a larger unit cell.

\subsection{Bulk properties}

Figure~\ref{MT_HC}(a) shows the temperature-dependent magnetic response $\chi(T)$ of {\Na} in an applied magnetic field of 1 T. The data was collected from field cooled powders with total mass of 0.3 grams. No evidence of magnetic ordering is observed in $\chi(T)$ down to 2 K, in contrast with a previous report on single crystals of {\Na} where a sharp antiferromagnetic transition was observed at $T_N = 30$ K~\cite{Wang2014}. No appreciable difference is observed between data collected under zero-field or field cooled conditions. Above 50 K, the magnetic susceptibility is weakly temperature dependent and is likely dominated by the Pauli term. To estimate the intrinsic susceptibility, $\chi_0$, we first fit the paramagnetic upturn below 50 K to a Curie-Weiss law and subtract it from the raw data. Next, we subtract the isotropic core diamagnetism~\cite{Bain2008} ($\chi_D \approx -5.02 \times 10^{-5}$ emu/mol). This gives an intrinsic $\chi_0 \sim 1.42(2)\times10^{-3}$ emu/mol Oe. The low temperature Curie-Weiss contribution is equivalent to approximately 10\% by mass of $S=1$ impurity, the origin of which we discuss later.

%No Curie-Weiss law fit could be performed in the high temperature region above 100 K.

Specific heat data shows no sign of a phase transition down to 1.9 K, as displayed in Figure~\ref{MT_HC}(b). Fitting the low temperature region (1.9 K $\leq T \leq 12$ K) to $C/T=\gamma + \beta T^2$ yields an electronic coefficient of $\gamma=11.7(2)$ mJ/Ru mol K$^2$ and a phononic contribution of $\beta=0.198(2)$ mJ/Ru mol K$^4$ (inset of Fig~\ref{MT_HC}(b)). The electronic coefficient is comparable to the values found for Sr$_2$RhO$_4$~\cite{Perry2006} and La-doped (Sr$_{1-x}$La$_x$)$_3$Ir$_2$O$_7$~\cite{Hogan2015, Hunter2015} ($x \geq 0.05$) and suggests that {\Na} may be a moderately correlated metal. The experimental values of $\chi_0$ and $\gamma$ gives a Wilson ratio of $R_W\approx8.9(1)$, which is much higher than the unity value for a free electron gas. The enhanced value of $R_W$ suggests that {\Na} is near an instability~\cite{Ikeda2000} and that it might be regarded as a moderately correlated metal with enhanced spin susceptibility.

%, $R_W=7.3 \times 10^4 \times \chi_0$(emu/mol)$/\gamma$(mJ/mol K$^2$),

%Within the Fermi-liquid theory, the Pauli susceptibility can be defined as $\chi_P=\mu_B^2N(E_F)$, where $\mu_B$ is the Bohr magneton constant and $N(E_F)$ is the electronic density of states at the Fermi energy ($E_F$). $N(E_F)$ can also be obtained from the Sommerfeld coefficient via $\gamma=(1/3)\pi^2k_B^2N(E_F)$, where $k_B$ is the Boltzmann constant. Using the experimental value of $\gamma$, the calculated Pauli susceptibility is $\chi_P=2.18(4) \times 10^{-4}$ emu/mol Oe, which is much smaller than the value of $\chi_0$ obtained from the magnetic susceptibility. This suggests that the spin susceptibility of the conduction electrons is strongly enhanced by exchange interactions, as it is observed for many metallic ruthenates~\cite{LeeS2006, Kimber2009, Kimber2010}.

\begin{figure*}
	\centering
	\includegraphics[width=1.9 \columnwidth]{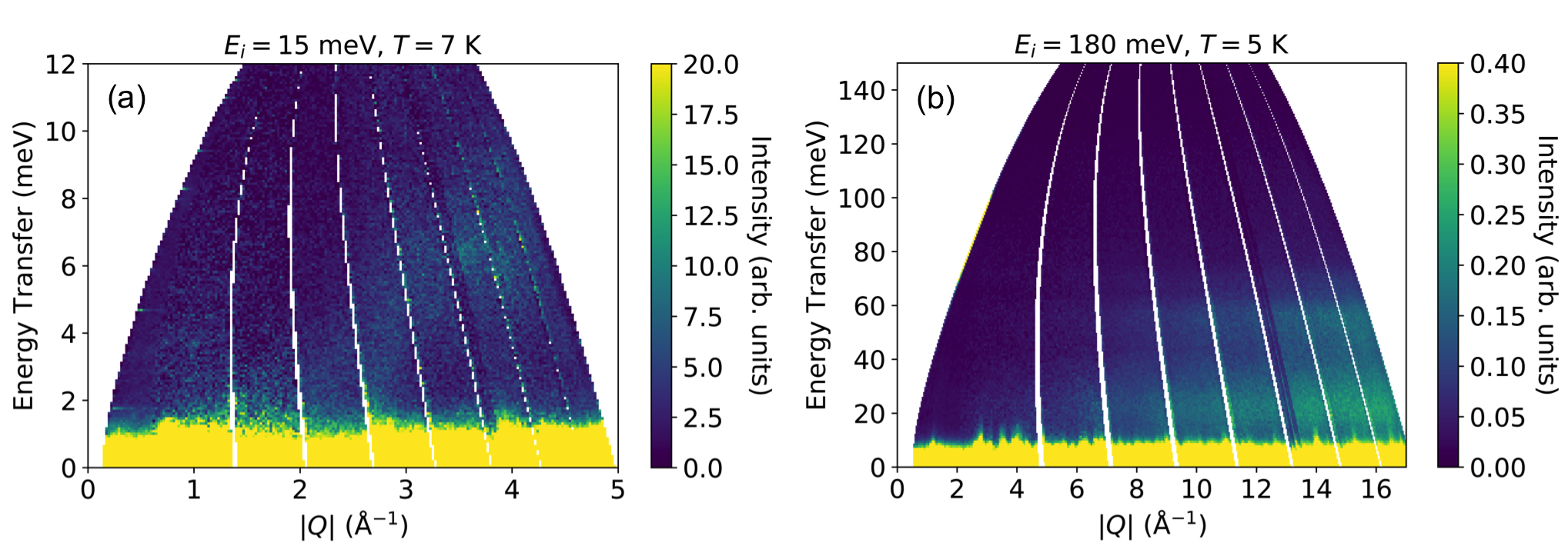}

	\caption{(Color online) INS spectra of {\Na} at low temperatures obtained using incident energies of (a) 15 meV and (b) 180 meV. No magnetic excitations are observed at any incident energy of the neutrons.}
	\label{INS}

\end{figure*}

No signatures of magnetic excitations of {\Na} were found in the inelastic neutron scattering measurements. The INS powder spectra collected at $T = 7$ K and 5 K with two different incident neutron energies, 15 meV and 180 meV, respectively, are shown in Figure~\ref{INS}(a-b). The color scale represents the observed intensity as a function of energy transfer and momentum transfer ($|Q|$). We note an absence of a continuum of spin excitations at low $|Q|$ and low energy transfer~\cite{Note1}. This behavior, characteristic of magnetic frustration, has been the most widely accepted evidence for a QSL state. The build-up of weak intensity at high-$|Q|$ region in Figure~\ref{INS}(b) comes from the phonon modes that become stronger at higher incident energies.

%The absence of long-range magnetic ordering is confirmed by the lack of Bragg peaks in the energy integrated Q-dependent spectra.

All the measurements mentioned above are consistent with our results from PES experiments. The inset of Figure~\ref{MT_HC}(a) shows the Fermi edges of polycrystalline Na$_2$RuO$_3$ at $T\leq50$ K with a photon energy of 6 eV. The PES measurements reveal a jump in the photoelectron intensity at the chemical potential, which is indicative of a small but finite density of states at the Fermi energy for all temperatures probed in these experiments. Similar PES experiments with a higher photon energy and correspondingly increased surface sensitivity did not detect any Fermi step. This strongly suggests that the PES data shown in the inset of Figure~\ref{MT_HC}(a) represent a finite bulk density of states at the chemical potential, consistent with the specific heat measurements. Although not performed in this present study, band structure calculations would be highly desirable to support the metallic nature of \Na{} and to confirm existing calculations available in the materials project database~\cite{Persson}.

\begin{figure}
	\centering
	\includegraphics[width=1 \columnwidth]{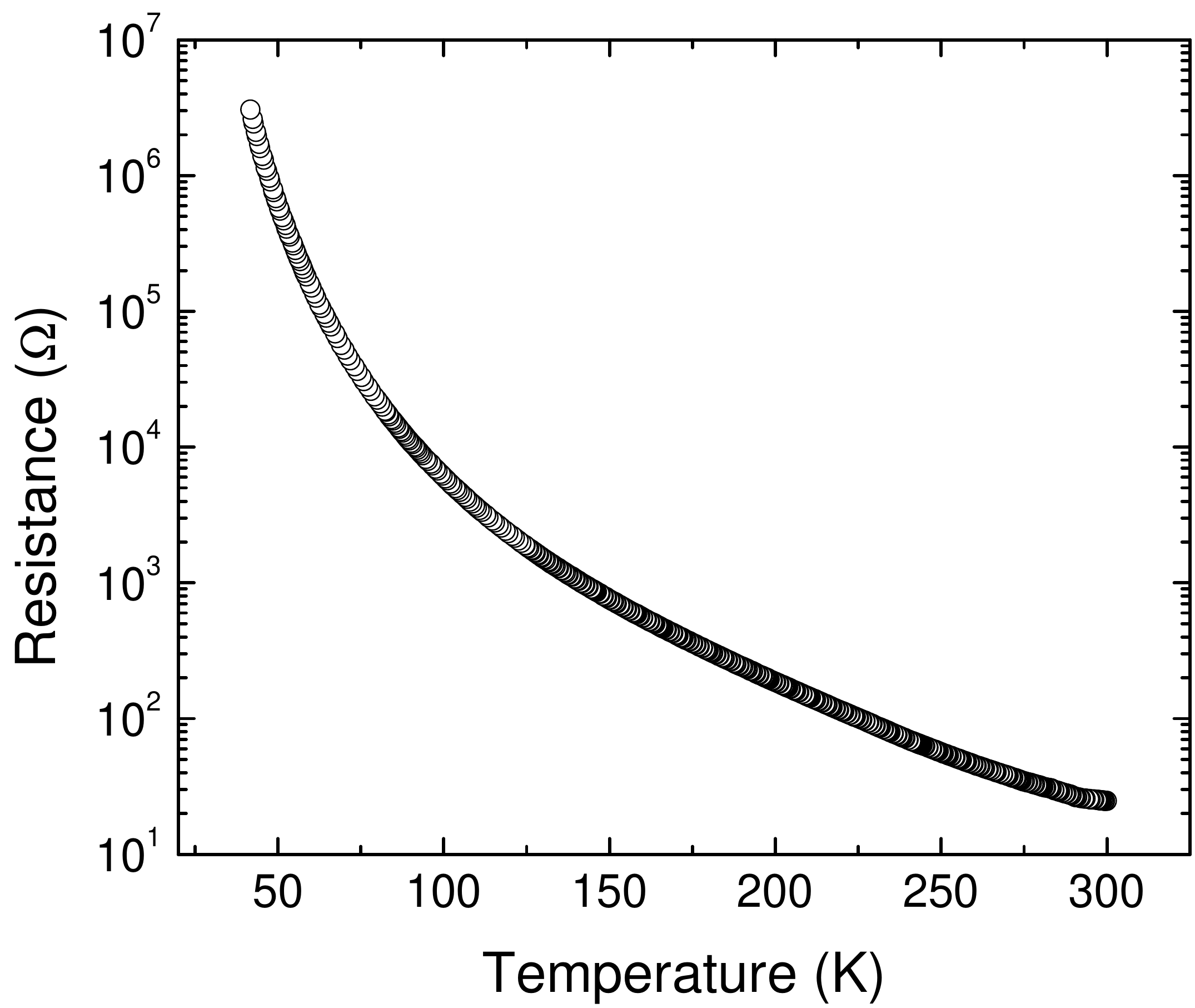}

	\caption{(Color online) Electrical resistance of polycrystalline Na$_2$RuO$_3$ as a function of temperature. }
	\label{R_PES}

\end{figure}

In contrast to all of our other data, the resistance was found to increase substantially on cooling indicating an insulating behaviour, as observed in Figure~\ref{R_PES}. These measurements were carried out using a conventional four-probe terminal method on a cold-pressed pellet, which was sintered in argon at 850 $^{\circ}$C for 1 hour to reduce the effects of grain boundaries. The annealing protocol did not degrade the sample, as confirmed by the X-ray diffraction measurements on the resulting pellets.

%All the measurements mentioned above are consistent with our results from PES experiments. The inset of Figure~\ref{MT_HC}(a) shows the Fermi edges of polycrystalline Na$_2$RuO$_3$ at $T\leq50$ K. The PES measurements reveals a jump in the photoelectron intensity at the chemical potential, which is indicative of a small but finite density of states at the Fermi energy. For all temperatures, a Fermi edge was observed in the PES spectra. 

%In order to confirm our supposition that this material is a moderately correlated Fermi liquid, we performed resistivity measurements, shown in Figure~\ref{R_PES}.
%We will discuss this discrepancy in the next section.

%While the temperature dependence of the resistance appears consistent with an insulating gap, 

%\subsection{Inelastic neutron scattering measurements}

%\subsection{Resistivity and PES measurements}

\section{Discussion}
\label{discussion}

The structural and chemical characterisation of our powders demonstrate that we have synthesised a nearly phase-pure compound of \textit{ordered} {\Na}. However, a complete solution of the  crystal structure is hindered by the presence of broad diffraction peaks, which are common for A$_2$MO$_3$ (A = Li, Na and M = Ir, Ru) with honeycomb ordered layers, and can be attributed to stacking disorder. Since in these layered materials the crystal grows perpendicular to the layer plane (here, along c-axis), the ordered honeycomb layers can nucleate in different stacking position, creating stacking faults. The deviation from the ideal sequence of honeycomb stacked layers leads to the peculiar triangular Warren-type peak shapes observed in Fig.~\ref{NPD}. While stacking faults can be modeled by Rietveld refinement using combined x-ray diffraction data sets~\cite{Bette2017, Bette2019, Bette2020, Diehl2018}, our attempts to tackle this problem proved unsuccessful mainly due to the complex background and intrinsic asymmetry of the peak shapes present in the TOF-NPD data. Nevertheless, our TOF-NPD analysis using the two-phase model reveals a high degree of ordering in the structure of our powders and suggests that the $C2/c$ crystal structure is a good candidate for the microstructural modelling of the stacking faults in {\Na}. Indeed, recent study on sodium-based batteries~\cite{Boisse2019} shows that modeling of stacking faults using similar space group ($C2/m$) is possible when using a combined X-ray diffraction data. Finally, the similarity of the phase percentage of the \textit{disordered} phase ($\approx$6\%) and the Curie-Weiss contribution to susceptibility at low temperature ($\approx$10\%) leads us to suggest that the \textit{disordered} phase might be responsible for the magnetic impurities. This is not unreasonable if the intra-plane Ru disorder causes the electrons to localise and contribute $S=1$ moment per Ru$^{+4}$ ion. The small amount of magnetic impurities may also be responsible for the enhanced Wilson ratio and can explain the absence of magnetic excitations in the INS data due to the proximity of a Stoner transition.

%focused on using two crystal structures present in literature, $C2/c$ (ordered) and $R\overline{3}mH$ (disordered) phases. Better refinements were achieved when adopting the two-phases approach used by Mogare et al.~\cite{Mogare2004}, which resulted in weight percentages of $\approx 93.93\%$ for $C2/c$ and $\approx 6.07\%$ for $R\overline{3}mH$ at 100 K, demonstrating the

Interestingly, our results reveal that {\Na} has a distinct ground state to that reported by Wang et. al.~\cite{Wang2014}. Single-crystals of {\Na} were claimed to order antiferromagnetically below $T_N=30$ K and to present a highly insulating ground state. We observe no evidence of magnetic order in our polycrystalline {\Na} in any bulk or spectroscopic probe: neither the NPD, nor bulk susceptibility show any feature at 30 K. Coincidentally, the features in the susceptibility from Ref~\onlinecite{Wang2014} are remarkably similar to those observed in Na$_3$RuO$_4$, whose structure consists of isolated tetramers of Ru$^{5+}$ ions ($S=3/2$) in a so-called lozenge configuration~\cite{Haraldsen2009, Regan2005}. We suggest that the single crystals from their study have been miscategorized as {\Na}. Remarkably, the structure of {\Na} is quite different to the iso-electronic sister compound Li$_2$RuO$_3$. In Li$_2$RuO$_3$, the polycrystalline low-temperature phase adopts a strongly distorted honeycomb lattice and dimerization, leading to the formation of a singlet ground state~\cite{Miura2007, Kimber2014, Park2016}, which is in contrast to {\Na} suggesting that the small ionic radius of lithium plays a crucial role in stabilising the singlet ground state.

%\textcolor{red}{One may be tempted to discuss the physical properties of \Na{} in terms of an excess of sodium since EDX measurements revealed that our samples are slightly Ru deficient. While sodium excess might be possible due to the flexibility of the structure as demonstrated by the batteries studies~\cite{Boisse2016, Boisse2019}, the high volatility of Na$_2$O at 950 $^{\circ}$C compared to RuO$_2$ suggests that the excess of sodium in the structure of \Na{} is unlikely to happen. It should also be noted that due to the low oxygen environment during the synthesis of \Na{} and the volatility of RuO$_2$, the ruthenium vacancies will likely be balanced by oxygen deficiencies to create impurity sites whereas no doping is realised. These impurity sites would act as scattering centers, increasing the resistivity of the material.}

The ground state of insulating ruthenates has been under debate in recent years due to the similarity of SOC, Hund's rule and crystal field energy scales. Traditionally,  perovskite ruthenates have been understood by arguing that the octahedral crystal field lifts the degeneracy of the $d$-orbital manifold and the four $4d$ electrons fill the $t_{2g}$ energy states to generate a low spin $S=1$ moment. However, if one considers the effect of the moderate SOC of the $4d$ electrons,  the $t_{2g}$ manifold could behave as a pseudo-$L_{\rm{eff}}=1$ (or $p$-state) generating a $J_{\rm{eff}}=0$ ground state with an excited state $J_{\rm{eff}}=1$ at an energy $\lambda$ (SOC energy scale) above the ground state, as postulated in \CaO ~\cite{Khaliullin2013,abragam_electron_2012}. Hence, one might expect a Van Vleck behaviour in the magnetism, where the mixing of the ground state and excited levels produces a temperature-independent contribution to the susceptibility that might explain the susceptibility of \Na{}. However, we believe that there are reasons to rule out this scenario in \Na{}. Firstly, the energy scale of the SOC in ruthenates is between 50 to 100 meV \cite{Tamai2019}; yet we do not observe any excited states in the INS data below 150 meV. Although we cannot rule out magnetic excitations at higher energy, which could be probed by resonant inelastic X-ray scattering. Secondly, a $J_{\rm{eff}}=0$ state with unquenched orbital angular momentum requires a high symmetry octahedral environment around the Ru$^{4+}$ ion but the \Na{} has a large octahedra distortion of 7\% [$r_l/r_s$, where $r_l$ ($r_s$) is the longest (shortest) Ru-O bonds], much greater than the Jahn-Teller distortions observed in Ca$_2$RuO$_4$ ($\approx 1.5\%$ at 10 K) \cite{Pincini2018a}. Such a large distortion is far more likely to quench the orbital magnetism within the $t_{2g}$ manifold and stabilise a configuration with a filled $d_{xy}$ orbital and half-filled $d_{yz,zx}$ orbitals, leading to an $S=1$ state in the localised limit.

Finally, the discrepancy between the bulk measurements in our polycrystalline \Na{} is perhaps most intriguing. Taken together, the susceptibility, heat capacity and the photoemission measurements indicate that this material has a Fermi liquid-like ground state. However, the insulating behaviour in the resistance measurements appears to contradict this hypothesis. We believe that the source of the discrepancy is the transport measurements made on our powder samples. It is a common practice in the community to determine the resistivity of powdered materials using cold or pressed samples and four-terminal d.c. voltage measurements. However, this can often lead to misleading results; resistance measurements on powdered samples contain contributions from two sources: intrinsic and grain boundary (extrinsic) and it is challenging to separate them. The electron hopping between grains can be modelled as a large energy barrier through which the electrons must tunnel, similar to electron motion in a scanning tunnelling microscope. This inter-grain hopping is the origin of the activation-like exponential increase in the resistance as the temperature is lowered and usually dominates over the intrinsic resistance of a grain. Moreover, the grain boundaries will introduce capacitance into the measurement circuit thus, ideally, the complex impedance should be measured as a function of frequency and temperature so that the real part of the intrinsic impedance can be extracted. For a good example, see Ref.~\onlinecite{Collier1989}, where the extracted grain boundary resistance in a pressed pellet of UO$_2$ had an activation-like temperature dependence with a barrier energy of ~0.13 eV and the grain boundary resistance was an order of magnitude larger than the observed intrinsic resistance. It is also worth pointing out that for transport measurements on powders, reproducibility is generally poor; each pressed pellet will have a unique grain boundary resistance due to a large number of factors, including porosity, grain stoichiometry, hygroscopicity and air sensitivity. In general, single crystals are the most reliable form on which to measure transport properties and measurements on powders should be treated with caution. For our system, if we neglect the resistivity data set, the simple conclusion is that \Na{} is a moderately correlated electron metal with modest magnetic enhancements, a conclusion supported by the photoemission spectroscopy experiments.

\section{Concluding remarks}

We have performed a detailed study of the synthesis and characterisation of polycrystalline {\Na} by using neutron diffraction, inelastic neutron scattering as well as susceptibility, heat capacity and photemission measurements. No sign of magnetic ordering or magnetic frustration is observed down to 1.5 K. Both magnetic susceptibility and heat capacity data indicate characteristics of significant electron correlation, with a large temperature independent Pauli paramagnetism and  moderate Sommerfeld coefficient. This is consistent with the observation of a Fermi edge in the PES measurements. In contrast, resistivity measurements show nonmetallic behaviour, which may be attributed to grain boundary effects. Thus, we classify {\Na} as a moderately correlated electron metal.

%Our results further demonstrate that resistivity measurements in powdered samples can lead to erroenous conclusions and highlight the importance of considering other probes to access a material's electronic properties.

\begin{acknowledgements}

We would like to thank G. Stenning and D. Nye for help with the instruments in the Materials Characterisation Laboratory at the ISIS Neutron and Muon Source. We also thank P. Manuel for help with the TOF-NPD measurements at WISH instrument and for useful comments and suggestions on our manuscript. Experiments at MARI instrument were supported by a beamtime allocation from the Science and Technology Facilities Council (STFC) under proposal RB1820455. We also acknowledge the use of the JEOL JSM-6610LV with Oxford Instruments EDX present in the Research Complex at Harwell (RCaH). This research was supported by the UK Engineering and Physical Sciences Research Council (EPSRC) (Grants No. EP/N027671/1 and No. EP/N034694/1). C.D.D. was supported by the EPSRC Centre for Doctoral Training in the Advanced Characterisation of Materials under Grant No. EP/L015277/1. F.B. and E.C. were supported by the Swiss National Science Foundation (SNSF). H.J. has received funding from the European Union’s Horizon 2020 research and innovation programme under the Marie Skłodowska-Curie Grant Agreement No. 701647.

\end{acknowledgements}

\bibliography{Bibliography}

\end{document}